\documentclass[twocolumn,showpacs,preprintnumbers]{revtex4}
\usepackage{dcolumn}
\usepackage{bm}
\usepackage{amsmath}
\usepackage{hyperref}
\usepackage[dvips]{graphicx}

\begin{document}

\title{Theory of current-driven motion of Skyrmions and spirals in helical magnets}
\author{Junichi Iwasaki$^{1}$}
\email{iwasaki@appi.t.u-tokyo.ac.jp}
\author{Masahito Mochizuki$^1$}
\author{Naoto Nagaosa$^{1,2}$}
\email{nagaosa@ap.t.u-tokyo.ac.jp}
\affiliation{$^1$ Department of Applied Physics, The University of Tokyo, 7-3-1, Hongo,
Bunkyo-ku, Tokyo 113-8656, Japan\\
$^2$Cross-Correlated Materials Research Group (CMRG), and Correlated
Electron Research Group (CERG), RIKEN-ASI, Wako, Saitama 351-0198, Japan
}
\date{\today}

\begin{abstract}
We study theoretically the dynamics of the spin textures, i.e., Skyrmion crystal (SkX) and 
spiral structure (SS), in two-dimensional helical magnets under external current.
By numerically solving the Landau-Lifshitz-Gilbert equation, it is found
that (i) the critical current density of the motion is much lower 
for SkX compared with SS in agreement with the recent experiment, 
(ii) there is no intrinsic pinning effect for SkX and the deformation 
of the internal structure of Skyrmion reduces the pinning effect dramatically, 
(iii) the Bragg intensity of SkX shows strong time-dependence as can be observed by 
neutron scattering experiment.
\end{abstract}

\maketitle
It has been recognized for a long time that the spin polarized electric 
current drives the motion of spin textures due to the spin transfer 
torque since the original theoretical proposal by ref.\cite{Slo,Berger}. 
The experimental demonstrations of this effect in the spin valves and
ferromagnetic domain wall systems followed~\cite{Maekawa}, 
which stimulated recent active researches on this phenomenon 
including those toward the applications such as 
racetrack memory~\cite{Parkin}. 
However, the threshold current density $j_c$  for the domain 
wall motion is rather high of the order of $10^{10}-10^{12}$ A/m$^2$~\cite{Maekawa}. 
Therefore, the Joule heating is a serious issue, and usually the experiments have been done by using 
the short pulses of electric current. There have been two origins of the pinning 
effect on the domain wall motion. One is the intrinsic pinning due to the 
magnetic anisotropy~\cite{Tatara}. More detailed analysis has showed that this intrinsic pinning 
occurs when only the Gilbert damping $\alpha$ is taken into account, and in the 
generic situation of the finite $\beta$ term there is no intrinsic pinning while the 
velocity is reduced~\cite{Tatara2}. When $\alpha=\beta$, there is completely no intrinsic 
pinning~\cite{Barnes}. Another origin is the extrinsic pinning due to the imperfections such as 
impurities and defects. In realistic situation, these two mechanisms are entangled 
and it is of vital importance to find some magnetic systems with lower critical 
current density. 

From this viewpoint, it is important to look for other types of magnetic 
textures for the current driven motion.  The Skyrmion~\cite{skyrme}, a topological spin 
texture where the spins point in all directions wrapping a sphere, is an 
interesting candidate in this respect. In magnets with non-centrosymmetric 
crystal structures, the Dzyaloshinskii-Moriya (DM) interaction~\cite{DM1,DM2} 
is allowed which naturally leads to the spiral spin structure. 
Under an external magnetic field $\vec{B}$, this spiral 
spins turns into the triangular crystal of Skyrmions in the narrow region in the $B-T$ phase diagram 
near the transition temperature $Tc$ in 3D as observed in neutron scattering experiments~\cite{Neutron}, 
and over a wide region of $B-T$ diagram in 2D as theoretically predicted~\cite{Yi2009}
and observed experimentally by Lorentz microscope~\cite{YuNature}. 
Remarkably, the small density of charge current ($j_c \sim 10^6$ A/m$^2$) 
is found to drive the motion of Skyrmions in 3D MnSi crystal as indicated by the 
slight rotation of the Bragg spot of the Skyrmion crystal in neutron scattering 
experiment~\cite{Current}. Furthermore, the current driven Skyrmion motion with even smaller 
$j_c \sim 10^5$ A/m$^2$ is concluded by the real space observation by Lorentz 
microscope in FeGe thin film~\cite{YuCurrent}. These observations have opened a new route to 
the manipulation of magnetic structures by ultra-low current density, 
and it is an important issue to study the mechanism of the pinning and 
the current-induced Skyrmions dynamics, which we undertake in the present letter.

The solid angle subtended by the spins in Skyrmion structure produces a fictitious 
magnetic field acting on the conduction electrons~\cite{Zang}, which gives rise to the 
topological Hall effect~\cite{TopHall1,TopHall2}. The motion of the 
Skyrmion produces the motion of this effective magnetic field, and hence the electromagnetic 
induction leading to an effective electric field in the transverse direction to the current. 
This results in a change in the Hall voltage when the Skyrmion motion sets in as predicted 
theoretically~\cite{Zang} and observed experimentally~\cite{PfHall}. 
This topological nature of Skyrmions also affects their dynamics since the coordinates, 
$X$ and $Y$, of the skyrmion core are canonical conjugate. 
Actually in ref.\cite{Zang}, the collective phonon modes and the pinning of the 
Skyrmion crystal have been studied, 
and a small threshold current density has been concluded. However, since the internal 
distortion of the Skyrmions has not been considered, it failed to explain the drastic 
difference between the pinning effect in spiral phase and that in Skyrmion phase, which was clearly 
shown in the experiment~\cite{YuCurrent}.  

\begin{figure*}[tb]
\begin{center}
\includegraphics[width=179mm]{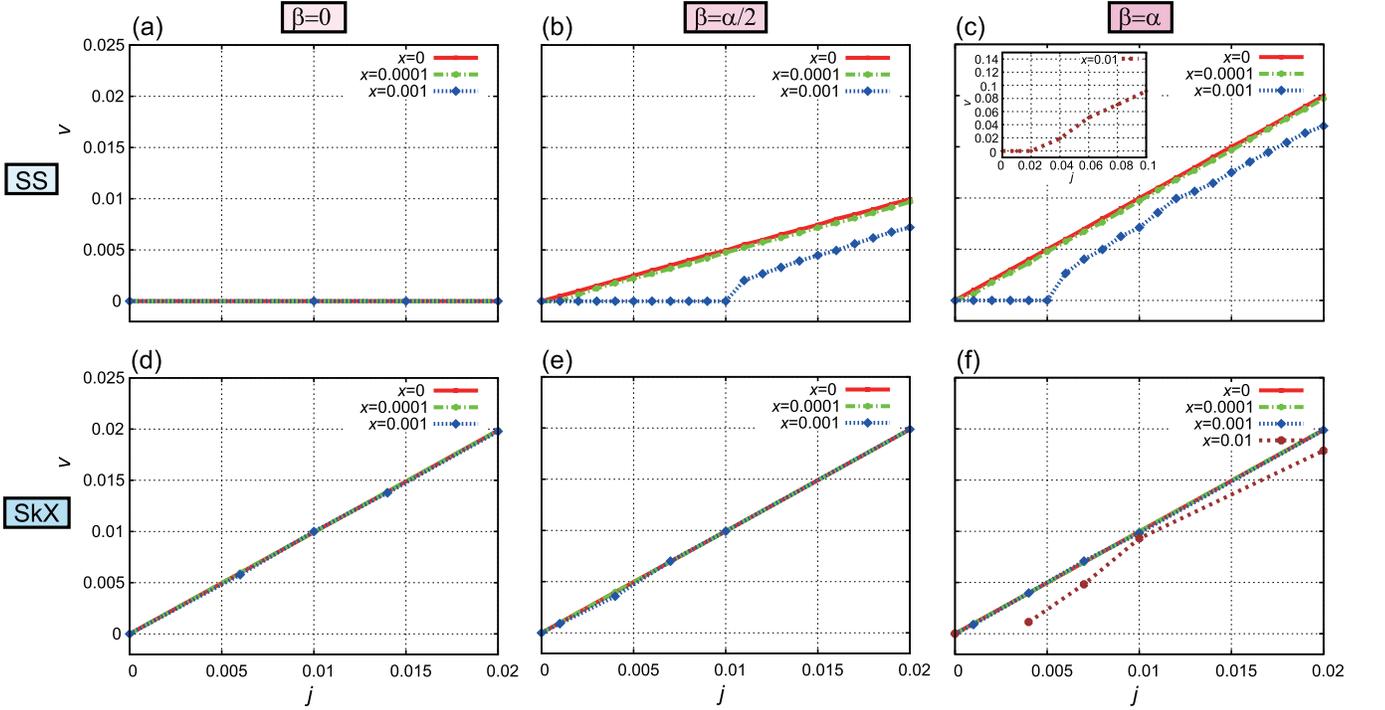}
\end{center}
\caption{
The relation between the current density $j$ and velocity $v$. 
Impurity density $x$ is $x=0,0.0001,0.001$ for $\beta=0,\alpha/2$, and $x=0,0.0001,0.001$ for $\beta=\alpha$.
(a) $\beta=0$ in SS. 
(b) $\beta=\alpha/2$ in SS. 
(c) $\beta=\alpha$ in SS. Note that the axis-ranges of the inset is different from that of outer one. 
(d) $\beta=0$ in SkX. 
(e) $\beta=\alpha/2$ in SkX. 
(f) $\beta=\alpha$ in SkX. 
}
\label{fig:1}
\end{figure*}

We start with the following classical Heisenberg Hamiltonian 
in a two-dimensional square lattice, which includes the ferromagnetic exchange 
interaction, DM interaction, Zeeman coupling with an external magnetic field, and magnetic 
anisotropy with an easy axis in a direction perpendicular to a lattice-plane due to 
the randomly-distributed imperfections: 
\begin{align}
\mathcal{H}=&-J\sum_{\vec{r}} \vec{M}_{\vec{r}} \cdot \left( \vec{M}_{\vec{r}+\vec{e_x}}+\vec{M}_{\vec{r}+\vec{e_y}} \right) \notag \\
&-D\sum_{\vec{r}} \left( \vec{M}_{\vec{r}} \times \vec{M}_{\vec{r}+\vec{e_x}} \cdot \vec{e_x} + \vec{M}_{\vec{r}} \times \vec{M}_{\vec{r}+\vec{e_y}} \cdot \vec{e_y} \right) \notag \\
&-\vec{B} \cdot \sum_{\vec{r}} \vec{M}_{\vec{r}} -A\sum_{\vec{r}\in I} M_z^2.
\end{align}
Here $I$ denotes the positions of impurities. The parameters are set as $J=1$, $D=0.18$, and $A=0.2$. The external magnetic field $\vec{B}$ is 
put as $\vec{B}=0.015 \ \vec{e_z}$ for the Skyrmion crystal, and $\vec{B}=0$ for the helical phase. 
We then study the spin dynamics at $T=0$ by numerically solving the Landau-Lifshitz-Gilbert equation 
including the terms which represent the coupling between spins and spin-polarized electric current density $\vec{j}$:
\begin{align}
\frac{{\rm d} \vec{M}_{\vec{r}}}{{\rm d} t}=& \gamma \vec{M}_{\vec{r}} \times B^{\rm eff}_{\vec{r}}
-\frac{\alpha}{M}\vec{M}_{\vec{r}} \times \frac{{\rm d} \vec{M}_{\vec{r}}}{{\rm d} t}
-\frac{pa^3}{2eM}\left( \vec{j} \cdot \vec{\nabla} \right) \vec{M}_{\vec{r}} \notag \\
&-\frac{pa^3\beta}{2eM^2} \left[ \vec{M}_{\vec{r}} \times \left( \vec{j} \cdot \vec{\nabla} \right) \vec{M}_{\vec{r}} \right],
\end{align}
where $B^{\rm eff}_{\vec{r}}=-\partial \mathcal{H}/\partial \vec{M}_{\vec{r}}$, $\gamma$ is the gyromagnetic ratio, $p$ is the spin-polarizability, and $a$ is the latttice constant. 
The Gilbert damping factor $\alpha$ is fixed to be $\alpha=0.04$, 
and three values for $\beta$ are considered, i.e., $\beta=0, \alpha/2, \alpha$.
We used the fourth-order Runge-Kutta method to solve this differential equation.
The size of the sample is  288$\times$288 and the periodic boundary condition is used. 
The initial spin configuration is obtained by the Monte-Carlo method and by further relaxing 
them at $j=0$ in the LLG simulation. After the sufficient convergence of the spin 
configuration, we switch on a steady electric current and observe the time evolution 
of the spins. For SS the direction of electric current is put to be parallel to the
wave vector of the spiral. In the following context, unit of time is $\tau \equiv \hbar/J$, and unit of current density is $\kappa \equiv \frac{2eSJ}{pa^2\hbar}$. 
$\tau$ and $\kappa$ are $\tau \simeq 6.5 \times 10^{-13}$ and $\kappa \simeq 2.0 \times 10^{13}$ for the typical set of parameters $J=1$ meV, $a=50$ nm and $p=0.1$. 

Figure 1 shows the the mean velocity $v$ as a function of the electric current density $j$ 
for SS and SkX with different values of $\beta$. With $\beta=0$,  
the SS cannot move even without the impurity or defect within the 
range of the current density of the present study as shown in Fig. 1(a). 
In sharp contrast, SkX moves freely with the velocity proportional to the 
current density for all values of the impurity concentration $x$ and $\beta$, 
even though a slight decrease of $v$ is observed for $x=0.01$ (Figs. 2(d)-(f)).
These results clearly indicate that the very strong intrinsic pinning exists 
for SS while no intrinsic pinning exists for SkX. 
The motion of the domain wall or SS is associated with 
the rotation of the spins within the plane, which is generated by the spin component perpendicular 
to them. Therefore, one needs the finite current density to compete with the 
magnetic anisotropy energy for the tilting of the spins away from the spin 
rotation plane~\cite{Tatara}. Usual intrinsic pinning comes from anisotropy with an easy plain. 
The system now considered also has an easy plain anisotropy because of DM interaction. 
The easy plane is perpendicular to the wave vector of the SS as determined by
the vector ${\vec D}$ of the DM interaction. 
On the other hand, there is no intrinsic pinning for SkX because of the
different canonical conjugate relation for the collective coordinates. Namely, 
the X and Y components of the center of mass motion of a Skyrmion
determines the equation of motion~\cite{Zang}, and hence uniform motion of the 
SkX does not require the distortion of the spin structure. 

\begin{figure}[tb]
\begin{center}
\includegraphics[width=85mm]{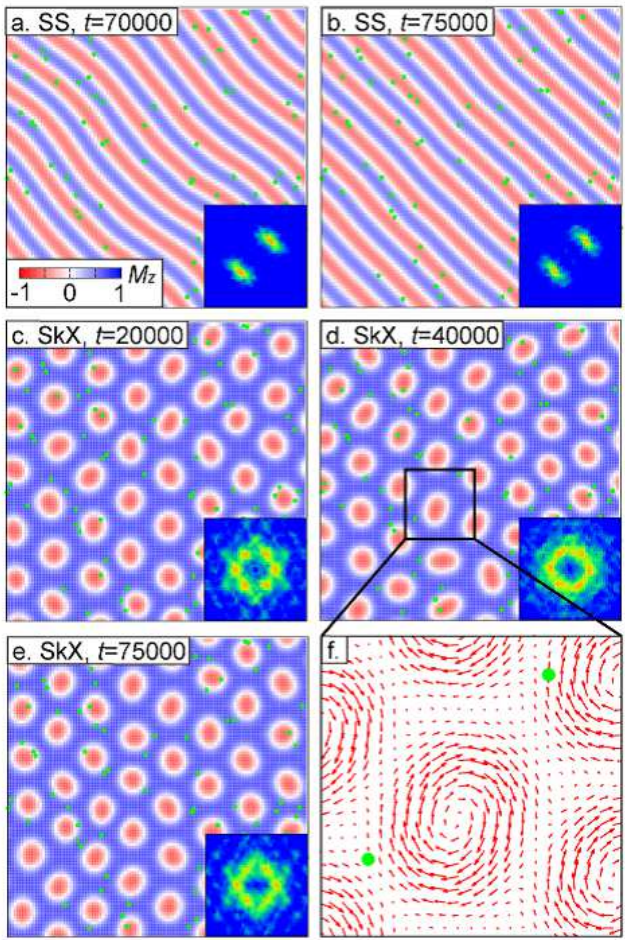}
\end{center}
\caption{
Snapshots of the moving spin textures at different times. 
Current density $j$ and impurity density $x$ are $j=0.006,x=0.001$ for SS, and $j=0.004,x=0.001$ for SkX. 
(a) SS, t=70000. (b) SS, t=75000. (c) SkX, t=20000. (d) SkX, t=40000. (e) SkX, t=75000. 
(f) The partial enlarged picture of (e).
}
\label{fig:2}
\end{figure}

For finite $\beta$, the intrinsic pinning does not give the finite threshold current density and 
especially in the case of $\beta=\alpha$, no intrinsic pinning is expected~\cite{Barnes},
where the pinning effect is purely the extrinsic one. 
In these cases, we can see that SS has a threshold current density $j_c$
which is a decrease function of $\beta$. This indicates that 
the interplay between the intrinsic and extrinsic mechanisms of the pinning occurs for SS. 
On the other hand, the velocity $v$ is hardly suppressed by the impurities 
for $x=0.0001, 0.001$ from the case of no impurity. 
Therefore, we can conclude that there is a large difference in the mechanism of the 
extrinsic pinning among these two phases, as discussed in depth below. 
From $x=0.01$ in Figs. 1(c) and (f), we can see that the threshold current density in 
SkX is at least factor 5 smaller than that in SS in the dense impurity case while
the ratio is much more in the dilute cases. 

In order to study the change in the spin configurations during the motion, we
take their snapshots in real space as shown in Fig. 2. 
Figures 2(a), (b) show the moving SS, where the position of each impurity is indicated by 
green dots. It is seen that even with $x=0.001$, the characteristic length scale of
the distortion is longer than the mean distance between impurities, and hence
the collective pinning mechanism applies here~\cite{pin}. Note that due to the 
one dimensional nature of the spiral the impurities are always acting on the
SS and distort it. This is not the case for SkX as shown in Figs. 2(c)-(f).
We can see that each Skyrmion avoids the impurities by deforming the triangular lattice 
and also transforming its shape. This is because the region between Skyrmions are
ferromagnetic state which does not feel the pinning, i.e., there is no energy
change with the translation of the ferromagnetic spin configuration.
This kind of motion is a key feature of SkX for explaining their ultra-low threshold current density. 
When the impurity works as an attractive center, the Skyrmion tends to rotate around it
rather than approaching to it. All of these features, i.e.,
the distortion of the triangular lattice and each Skyrmion, and
the rotating motion around the impurities, work together to
reduce the pinning effect for SkX.

Deformation of SkX explained above should be able to be detected
experimentally. For that purpose, we propose the neutron scattering
experiment on the time-dependence of the Bragg peaks of the SkX. 
Figure 3 shows the change in the peak value of the spin structure 
function $S({\vec k})$ at the Bragg points for the SkX.
The values are normalized so that the integral of $S({\vec k})$ over
the first Brillouine zone is 1. Because of the uniform magnetization 
under the external magnetic field, the component at ${\vec k}= (0, 0)$
is finite, and hence the red line for $x=0$ corresponds to 
the perfect SkX. 
For finite $x$, the peak value fluctuates in time. 
This means that there is the time-dependent deformation of SkX.
When we take the Fourier transformation (Fig. 3), 
there are two characteristic frequencies. 
One is around $\omega=0.045\pi(=0.1413)$, which corresponds to the 
periodic distortion and recovery of SkX synchronized with its 
motion with one lattice constant of SkX. It is because when a Skyrmion approaches 
imperfections the triangular lattice is little broken, but soon after the Skyrmion
passed by imperfections the lattice begins to return to triangular lattice. 
The other lower one $\omega<0.005\pi(=0.0157)$ corresponds to 
the slow and collective distortion of SkX. 
\begin{figure}[tb]
\begin{center}
\includegraphics[width=86mm]{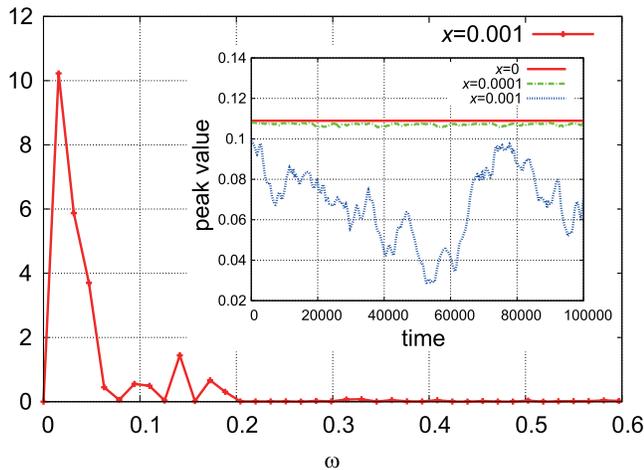}
\end{center}
\caption{ 
The inset of the figure is the time-dependence of the Bragg peak value in SkX ($j=0.004$). 
The outer figure is the spectra of $x=0.001$.
}
\label{fig:3}
\end{figure}

In summary, we have theoretically studied the dynamics of 
spiral spin structure (SS) and Skyrmion crystal (SkX) in helical magnets with  
impurities or defects driven by electric currents. We have confirmed the strong intrinsic 
pinning effect for SS, while no intrinsic pinning for SkX. 
Also the extrinsic pinning effect is found to be much weaker for SkX compared with SS.
By the real-space snapshots, the origin of this difference is revealed;
SkX avoid the imperfections by the deformation of the triangular lattice 
and also the shape of each Skyrmion together with the rotating motion around the impurity.
This leads to the very weak threshold current density for SkX as observed
experimentally compared with SS. Finally, we have proposed the neutron scattering experiment 
to detect the time-dependent Bragg peaks showing the periodic change in the 
SkX. 

During the completion of this paper we became aware of a recent
relevant paper by Liu and Li~\cite{condmat}, 
which addresses the pinning mechanism of
Skyrmions in chiral magnets. They consider the one and two Skyrmions 
with one pinning center, while the present work focuses on the 
crystal of Skyrmions with many pinning centers. 
We are grateful for the insightful discussions with Prof. Y.Tokura and 
Dr. X.Z. Yu. This work is supported by Grant-in-Aids for Scientific
Research (No. 24224009)
from the Ministry of Education, Culture, Sports, Science and
Technology of Japan, and also by Funding Program for World-Leading
Innovative R\&D on Science and Technology (FIRST Program).
MM was supported by G-COE Program ``Physical Sciences Frontier" from
MEXT Japan.

\end{document}